\documentclass[doublecol]{epl2} 

\title{On the efficiency at maximum cooling power}

\author{Y. Apertet\inst{1} \and H. Ouerdane\inst{2}\inst{3} \and A. Michot\inst{2} \and C. Goupil\inst{2}\inst{3} \and Ph. Lecoeur\inst{1}}
\shortauthor{Y. Apertet \etal}

\institute{                    
  \inst{1} Institut d'Electronique Fondamentale, Universit\'e Paris-Sud, CNRS, UMR 8622, F-91405 Orsay, France, EU\\
  \inst{2} Laboratoire CRISMAT, UMR 6508 CNRS, ENSICAEN et Universit\'e de Caen Basse Normandie, 6 Boulevard Mar\'echal Juin, F-14050 Caen, France, EU\\
  \inst{3} Universit\'e Paris Diderot, Sorbonne Paris Cit\'e, Institut des Energies de Demain (IED) URD 0001, 75205 Paris, France, EU\\
}


\abstract{The efficiency at maximum power (EMP) of heat engines operating as generators is one corner stone of finite-time thermodynamics, the Curzon-Ahlborn efficiency $\eta_{\rm CA}$ being considered as a universal upper bound. Yet, no valid counterpart to $\eta_{\rm CA}$ has been derived for the efficiency at maximum \emph{cooling} power (EMCP) for heat engines operating as refrigerators. In this Letter we analyse the reasons of the failure to obtain such a bound and we demonstrate that, despite the introduction of several optimisation criteria, the maximum cooling power condition should be considered as the genuine equivalent of maximum power condition in the finite-time thermodynamics frame. We then propose and discuss an analytic expression for the EMCP in the specific case of exoreversible refrigerators.}

\begin{document}

\maketitle

\section{Introduction}
The study of energy conversion has acquired the status of science with the theoretical derivation by Carnot of an upper bound for the efficiency of heat engines, the so-called Carnot efficiency \cite{Carnot1824}. Though the far-reaching implications of Carnot's result were not fully apprehended at the time of its derivation, it constitutes the first formulation of the second law of thermodynamics. In Carnot's model, the maximum efficiency is reached for an infinite thermodynamic cycle duration and the output power consequently vanishes. Further, since Carnot's system is supposed to be perfect, none of its constituents may ensure causality so no description of the time evolution of the system is possible. Hence, although practical considerations were the motivation of Carnot's work, his results do not apply to actual engines.

The advent of finite-time thermodynamics (FTT) in the 1950's permitted a drastic improvement of thermodynamic analyses of power generation systems: the newly built atomic power plants were designed to produce as much power as possible so that the energy conversion efficiency no longer was the quantity to maximise at all cost, albeit it is desirable to maintain it as high as possible when the system works at maximum output power. In the first works, causality was only ensured by the introduction of dissipative elements between the perfect (Carnot) engine and the two thermal reservoirs, at temperatures $T_{\rm hot}$ and $T_{\rm cold}$ ($T_{\rm hot} > T_{\rm cold}$): the model system thus defined is \emph{endoreversible}.

Yvon \cite{Yvon1955} and Novikov \cite{Novikov1957, Novikov1958} independently derived a general expression for the efficiency at maximum power. This expression, independent of the particulars of the studied system, was however labeled according to two other authors, Curzon and Ahlborn, who rederived it in an elegant way in the 1970's \cite{Curzon1975}. So the so-called Curzon-Ahlborn efficiency $\eta_{\rm CA}$ has since then been the touchstone of FTT:
\begin{equation}
\eta_{\rm CA} = 1 - \sqrt{\frac{T_{\rm cold}}{T_{\rm hot}}} = 1 - \sqrt{1 - \eta_{\rm C}}
\end{equation}
\noindent where $\eta_{\rm C} = 1- T_{\rm cold}/T_{\rm hot}$ is the Carnot efficiency. General discussions on the EMP may be found in Refs.~\cite{ReviewFTT1} and \cite{ReviewFTT2}.

A quite different expression for the EMP was recently obtained by Schmiedl and Seifert \cite{Schmiedl2008}:
\begin{equation}
\eta_{\rm SS} = \frac{\eta_{\rm C}}{2 - \gamma \eta_{\rm C}}
\end{equation} 
\noindent where $\gamma$ is a parameter comprised between 0 and 1. We demonstrated in Ref.~\cite{Apertet2012PRE1} that this expression actually corresponds to a qualitatively different class of heat engine models: the thermal contacts between the thermal reservoirs and the system are supposed to be perfect and the irreversibility necessary to ensure causality is only provided by internal dissipations such as, e.g., frictions, Joule heating. Such model systems are \emph{exoreversible}.

\begin{figure}
	\centering
		\includegraphics[width=0.45\textwidth]{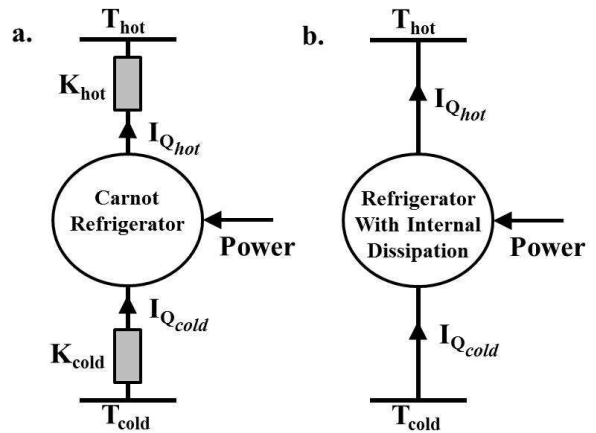}
	\caption{Thermodynamic picture of two classes of refrigerators: a. endoreversible refrigerators, b. exoreversible refrigerators.}
	\label{fig:figure1}
\end{figure}

Among the possible sources of irreversibilities, besides the internal dissipations and the finiteness of the thermal coupling to the heat reservoirs, there are heat leaks, i.e. heat that flows directly from the hot to the cold reservoir without taking part to the energy conversion process. This third source of irreversibility cannot ensure causality \cite{Apertet2012PRE1}; and, since heat leaks are detrimental to the energy conversion efficiency and have no bearing on causality, these are excluded from both the endoreversible and the exoreversible models. Neglecting heat leaks amounts to making the strong coupling assumption \cite{{VandenBroeck2005}}, which is mandatory to obtain the efficiencies $\eta_{\rm CA}$ and $\eta_{\rm SS}$. We thus assume that the refrigerators operate in the strong coupling regime. These are represented in Figs.~\ref{fig:figure1}.a and ~\ref{fig:figure1}.b, in the endoreversible and exoreversible configurations, respectively.

While the FTT description of engines working as generators benefits from well defined efficiencies such as $\eta_{\rm CA}$ and $\eta_{\rm SS}$, things are not so clear for refrigerators: the search for a simple expression of the efficiency at maximum cooling power (EMCP), i.e., when the heat current coming from the cold reservoir (the cooling power) is maximum, remains unsuccessful.
Although several criteria have been proposed as equivalents to the CA efficiency \cite{Agrawal1990,Yan1990,Velasco1997,Allahverdyan2010,deTomas2012,Luo2013}, none of them can be considered as a true equivalent. In this article, using the example of a thermoelectric module, we discuss the reasons for such a failure to find the counterpart of $\eta_{\rm CA}$ and we propose an expression for the EMCP when the refrigerator is described by an exoreversible model.

Our article is organised as follows. In section 2, we present briefly the specific model of a thermoelectric cooler on which we base our reasoning. In section 3, we discuss the facts that preclude the derivation of a Curzon-Ahlborn analogue to the EMCP in the refrigerator regime. In Section 4, we derive a general expression for the EMCP when the refrigerator is exoreversible. We end the paper with a discussion and concluding remarks.

\section{Thermoelectric model}
We use a thermoelectric module as a toy model to discuss the EMCP. This particular model system is valuable in the sense that it is sufficiently broad to lead to conclusions that may easily be generalised, e.g. see \cite{Gordon1991, Apertet2012PRE1}, and that it also is more apprehensible than pure formalism such as the general Onsager force-flux formalism used in Refs.~\cite{VandenBroeck2005} and \cite{Izumida2012}. In particular, the notion of internal dissipation, i.e., internal heat production resulting from the energy conversion process, at the heart of the present article, appears naturally in the study of thermoelectric refrigerators with the introduction of an internal electrical resistance $R$.

\begin{figure}
	\centering
		\includegraphics[width=0.48\textwidth]{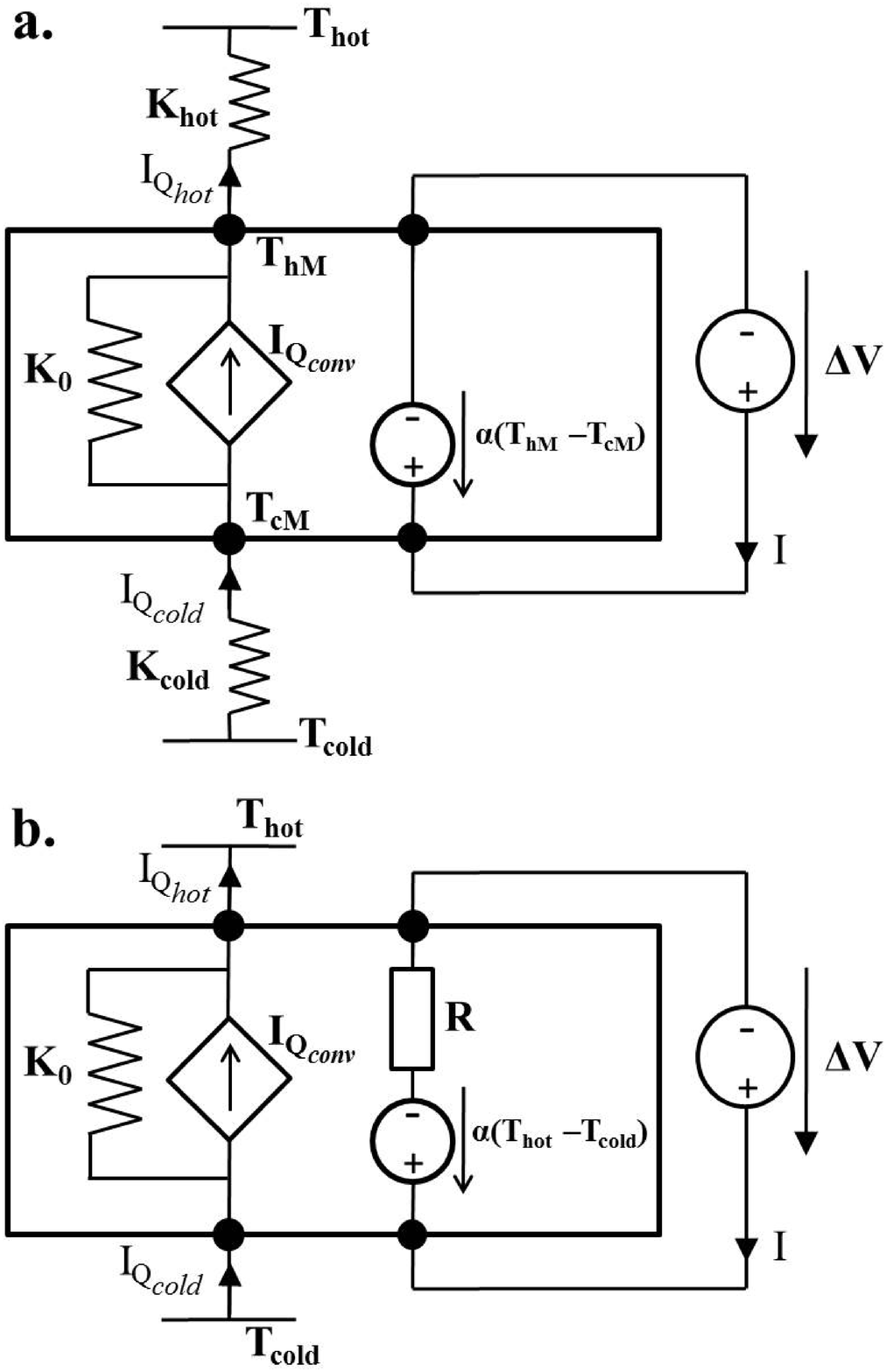}
	\caption{Description of the thermoelectric refrigerator. a: Endoreversible refrigerator, b: Exoreversible refrigerator. Since strong coupling is assumed, the thermal conductance under open electrical circuit $K_0$ vanishes. The controlled current source delivering ${I_Q}_{\rm conv}$ represents the convective contribution to the thermal flux \cite{Apertet2012JPCS}.}
	\label{fig:figure2}
\end{figure}

We start our study with the general model presented in Ref.~\cite{Apertet2012EPL} where both internal dissipation and finite thermal coupling to the reservoirs are considered. However, we solely focus here on the two extreme cases corresponding to endoreversible and exoreversible refrigerators, represented in Figs.~\ref{fig:figure2}.a and \ref{fig:figure2}.b respectively. A \emph{by-pass} thermal conductance $K_0$ is shown in both circuit representations; but we choose to neglect it in the present work since accounting for this quantity brings nothing of interest in our finite-time thermodynamics analysis of refrigerators: though heat leaks are of practical interest, they may be viewed as pure \emph{parasitic} processes that lower the overall system's performance even though they do not pertain to energy conversion. Setting $K_0 = 0$ is justified by the strong coupling assumption, and it simplifies the calculations and hence the analyses of the fundamental properties of the systems under consideration.

To determine the working conditions for maximum cooling, the knowledge of heat fluxes on the hot and cold sides of the thermoelectric module is required. For the endoreversible model, the thermal flux $I_{Q}$ inside the thermoelectric cooler results from the \emph{convective} process only \cite{Apertet2012JPCS} and is thus proportional to the electrical current $I$: ${I_{Q} = {I_Q}_{\rm conv} = \alpha T I}$ where $\alpha$ is the Seebeck coefficient and $T$ is the local temperature. Due to the finite thermal conductances of the heat exchangers, $K_{\rm hot}$ and $K_{\rm cold}$, the temperatures on each side differ from $T_{\rm hot}$ and $T_{\rm cold}$; these are denoted $T_{\rm hM}$ and $T_{\rm cM}$. Continuity of the thermal flux at the interfaces yields the following equations for the thermal fluxes on the hot and cold sides:
\begin{equation}\label{thermalflux}
\begin{array}{c}
I_{Q_{\rm hot}} = \alpha T_{\rm hM} I = K_{\rm hot} (T_{\rm hM} - T_{\rm hot})\\
~\\
I_{Q_{\rm cold}} = \alpha T_{\rm cM} I = - K_{\rm cold} (T_{\rm cM} - T_{\rm cold})
\end{array}
\end{equation}
\noindent since the flux thermal in the heat exchangers is assumed to follow Fourier's law for heat conduction. The ensuing expressions of the temperatures $T_{\rm hM}$ and $T_{\rm cM}$
\begin{equation}\label{temp}
\begin{array}{c}
T_{\rm hM} = \frac{\displaystyle K_{\rm hot} T_{\rm hot}}{\displaystyle K_{\rm hot} - \alpha I}~~{\rm and}~~T_{\rm cM} = \frac{\displaystyle K_{\rm cold} T_{\rm cold}}{\displaystyle K_{\rm cold} + \alpha I}
\end{array}
\end{equation}
\noindent show their explicit dependence on the electrical current $I$. Inserting these latter in Eq.~(\ref{thermalflux}) yields:
\begin{equation}
I_{Q_{\rm hot}} =  \frac{\displaystyle \alpha K_{\rm hot} T_{\rm hot} I}{\displaystyle K_{\rm hot} - \alpha I}~~{\rm and}~~I_{Q_{\rm cold}} = \frac{\displaystyle \alpha K_{\rm cold} T_{\rm cold} I}{\displaystyle K_{\rm cold} + \alpha I}
\end{equation}

For the exoreversible models, the temperatures on each side are well defined by the thermal reservoirs but additional terms related to internal dissipation, i.e., Joule heating, must be considered:
\begin{equation}\label{fluxexo}
\begin{array}{c}
I_{Q_{\rm hot}} =  \alpha T_{\rm hot} I  + \frac{\displaystyle 1}{\displaystyle 2}R I^2\\
~\\
I_{Q_{\rm cold}} = \alpha T_{\rm cold} I  - \frac{\displaystyle 1}{\displaystyle 2}R I^2
\end{array}
\end{equation}

\noindent Half of the dissipated heat is rejected to the hot reservoir, and half is rejected to the cold side \cite{Ioffe}.

\begin{figure}
	\centering
		\includegraphics[width=0.48\textwidth]{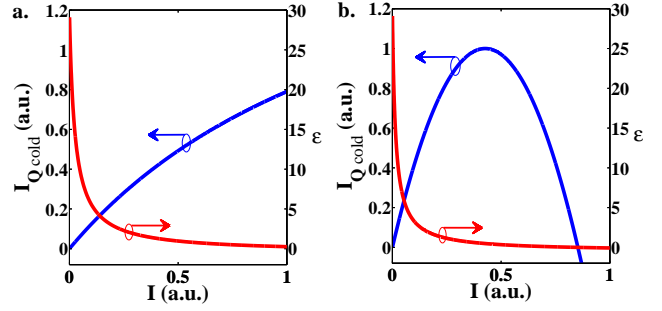}
	\caption{Cooling power scaled to its maximum value and cooling efficiency against the electrical current: a. endoreversible refrigerator, b. exoreversible refrigerator.}
	\label{fig:figure3}
\end{figure}

Now, the cooling efficiency reads:
\begin{equation}
\epsilon = \frac{I_{Q_{\rm cold}}}{P}  = \frac{I_{Q_{\rm cold}}}{I_{Q_{\rm hot}} - I_{Q_{\rm cold}}} 
\end{equation}
\noindent for both the endoreversible and exoreversible models. The cooling efficiency $\epsilon$ and cooling power $I_{Q_{\rm cold}}$ are plotted against the electrical current $I$ on Figure~\ref{fig:figure3}. We notice that the efficiency follows the same trend for both models. However, the behaviour of the cooling power strongly depends on the assumption of endo- or exoreversibility: for endoreversible engines (Fig.~\ref{fig:figure3}.a) it monotonically increases while for exoreversible engines (Fig.~\ref{fig:figure3}.b), it has a maximum. Indeed in the latter case, as $I$ increases both the thermoelectric convective heat flux and Joule heating increase, and when the dissipated heat becomes preponderant over the transported heat, the net cooling power decreases, and even becomes negative for high electrical currents. Figure~\ref{fig:figure3} thus illustrates the main difference between endoreversible and exoreversible models. In the next sections, we discuss the notion of efficiency at maximum cooling power using this example. 

\section{EMCP for endoreversible engines}
We now present and discuss the validity of various propositions to obtain an efficiency at maximum cooling power in the refrigerator regime, analogue to the Curzon and Ahlborn efficiency for generators.

\subsection{Looking for a maximum}
The main problem here with the endoreversible engine is that if a maximum exists for the power in the generator regime, there is no such maximum for cooling power in the refrigerator regime as illustrated on Figure~\ref{fig:figure3}. To obtain a maximum, different authors used various hypotheses. While Agrawal and Menon \cite{Agrawal1990} considered the contribution of the adiabatic steps (neglected in the derivation of the CA efficiency) to the total cycle time using the molar heat capacity at constant volume of the working substance, Velasco and co-workers \cite{Velasco1997} did not use the cooling power as the quantity to optimize but rather the per-unit-time cooling efficiency, i.e., $\epsilon / t$ where $t$ is the overall thermodynamic cycle time. It is interesting to note that both in Refs.~\cite{Agrawal1990} and \cite{Velasco1997}, the changes introduced to derive the optimal performance in the refrigerator regime were also applied to the engine in generator regime. Velasco and co-workers found that the CA efficiency also corresponds to the efficiency at maximum per-unit-time efficiency. Conversely, Agrawal and Menon found that $\eta_{\rm CA}$ is no longer the reference for heat engines when the adiabatic steps are taken into account.

Also noticing the lack of maximum for the cooling power of endoreversible engine, Yan and Chen \cite{Yan1990} proposed to use a criterion based on both cooling efficiency and cooling power: $\epsilon I_{Q_{\rm cold}}$. This new criterion was then adopted by other authors in Refs.~\cite{Allahverdyan2010, Luo2013, deTomas2012}. However, Yan and Chen only made the change of performance coefficient for the refrigerator: $\eta_{\rm CA}$ was kept as the reference for heat engine efficiency although the optimization was realised with a distinct criterion. Recently de Tom\'as and co-workers \cite{deTomas2012} suggested that these different optimisation strategies might be unified in a single criterion denoted $\chi$ and defined as:
\begin{equation}
\chi = \frac{z {Q_{\rm in}}}{t_{\rm cycle}}
\end{equation}
\noindent with $z$ being the efficiency, $Q_{\rm in}$ being the heat absorbed by the engine, and $t_{\rm cycle}$ being the time of the thermodynamic cycle.

\subsection{On the coherence of the $\chi$ criterion}
This $\chi$ criterion leads to two different expressions for refrigerators and generators when replacing $z$ and $Q_{\rm in}$ by the suitable values for each case. For a generator, the criterion becomes:
\begin{equation}
\chi^{(\rm G)} = \frac{\eta {Q_{\rm hot}}}{t_{\rm cycle}} = \frac{W}{t_{\rm cycle}},
\end{equation}
\noindent and permits to recover the output power used in the CA derivation, while for a refrigerator one gets:
\begin{equation}
\chi^{(\rm R)} = \frac{\epsilon {Q_{\rm cold}}}{t_{\rm cycle}},
\end{equation}
\noindent which corresponds to the criterion defined by Yan and Chen \cite{Yan1990}. An interesting feature of the efficiency at $\chi_{\rm max}$ is the independence regarding particular engine properties: it was already the case for $\eta_{\rm CA}$ but it is also true for the expression obtained for refrigerators with strong coupling and left-right symmetry:
\begin{equation}
\epsilon_{\chi_{\rm max}} = \sqrt{1 + \epsilon_{\rm C} } - 1.
\end{equation}
\noindent where $\epsilon_{\rm C} = T_{\rm cold} / (T_{\rm hot} -T_{\rm cold})$ is the Carnot efficiency for refrigerators. The above expression has been extended to non-symmetric refrigerators in Refs.~\cite{Izumida2012} and \cite{Wang2012}.

However, in our opinion, the $\chi$ criterion is an elegant but artificial way to conciliate the optimal coefficient of performance defined by Yan and Chen for refrigerators with the classical derivation of Curzon and Ahlborn. The relevance of such criterion must be questioned: if the measure of performance has to involve the power ($P$ or $I_{Q_{\rm cold}}$) and the efficiency ($\eta$ or $\epsilon$), the quantity corresponding to the product $\epsilon I_{Q_{\rm cold}}$ for generators should not be $P$ but rather the product $\eta P$, a criterion proposed by Stucki \cite{Stucki1980}. The choice of maximum power as the coefficient of performance in FTT is motivated by the fact that both power and efficiency are desirable for heat engines but a compromise should be made between these two quantities; the working conditions for maximum efficiency and maximum power thus correspond to bounds delimiting a range of possible working conditions satisfying this compromise \cite{Salamon2001}. The criterion $\eta P$ is therefore closer to the $\Omega$-criterion defined by Calvo Hern\'andez and coworkers \cite{Calvo2001}(applied to refrigerators in Refs.~\cite{Sanchez2010} and \cite{deTomas2013}) or to the ecological optimisation criterion defined by Angulo-Brown \cite{Angulo1991}, than to a power maximisation. Choosing one of these criteria instead of maximum cooling power indeed leads to focus not on the bound of this range but rather on a particular realisation of the efficiency/power trade-off. We thus come to state that in order to be equivalent to a Curzon-and-Ahlborn-like derivation, the criterion chosen to optimise refrigerators in the frame of the FTT, should also contain this idea of desirable limit for working conditions: this condition is met only for maximum cooling power. With these considerations in mind, $\epsilon_{\chi_{\rm max}}$ should not be considered as the genuine counterpart of the Curzon-Ahlborn efficiency and we assert that such a counterpart does \emph{not} exist because of the absence of a cooling power maximum for endoreversible systems.

\section{EMCP for exoreversible engines}
Contrary to the endoreversible configuration, a maximum for cooling power exists for exoreversible refrigerators as shown in Figure~\ref{fig:figure3}.b; so there is no need to define a new criterion in this case. We now derive an expression for the EMCP for a thermoelectric module, which we generalise and discuss.

\subsection{Case of a thermoelectric cooler}
With Eq.~(\ref{fluxexo}), we see that the electrical current maximising the cooling power is ${I_{\rm MCP} =  \alpha T_{\rm cold} / R}$ so the corresponding cooling power reads:
\begin{equation}
I_{Q_{\rm cold}}^{(\rm max)} = \frac{\alpha^2 T_{\rm cold}^2 }{2R}.
\end{equation}
\noindent Note that this maximum cooling power decreases as the internal dissipations, embodied in the internal electrical resistance $R$, increase. Now, using the general expression for the cooling efficiency:
\begin{equation}
\epsilon = \frac{I_{Q_{\rm cold}}}{P} = \frac{\alpha T_{\rm cold} I - RI^2/2}{\alpha \Delta T I + RI^2},
\end{equation}
\noindent we derive the EMCP for the thermoelectric refrigerator:
\begin{equation}
\epsilon_{\rm MCP} = \frac{\epsilon_{\rm C}}{2( 1 + \epsilon_{\rm C})}
\end{equation}
This expression does not depend on particular values of the engine's characteristics, i.e., $\alpha$ and $R$, but only on the thermal reservoir temperatures $T_{\rm hot}$ and $T_{\rm cold}$ just as $\eta_{\rm CA}$ does. We do not claim however that this is the counterpart of the Curzon and Ahlborn efficiency; this expression looks rather like Schmiedl-Seifert efficiency $\eta_{\rm SS}$ which was obtained using an exoreversible model system too. Yet, the efficiency $\eta_{\rm SS}$ contains a $\gamma$ factor that does not appear in the considered thermoelectric example. In order to derive a complete equivalent to $\eta_{\rm SS}$ for refrigerators, we make use a of generalised model.
 
\subsection{Generalisation}
We recently demonstrated that the $\gamma$ parameter for autonomous engines, such as thermoelectric modules, may be interpreted as the fraction of the dissipated heat inside the engine that is released into the hot thermal reservoir \cite{Apertet2013}. The dissipated heat that flows back to the hot reservoir may be recycled for energy conversion since it is available again to fuel the engine. Conversely, all the heat released to the cold reservoir is lost for good. This explains why increasing $\gamma$ yields an increase of $\eta_{\rm SS}$. For thermoelectric systems, $\gamma = 1/2$, but this parameter might be different for other types of engines, in which case Eq.~(\ref{fluxexo}) becomes:
\begin{equation}
\begin{array}{c}
I_{Q_{\rm hot}} = \alpha T_{\rm hot} I  + \gamma R I^2 \\
~\\
I_{Q_{\rm cold}} = \alpha T_{\rm cold} I  - (1-\gamma)R I^2 \\
\end{array}
\end{equation}
\noindent with $\gamma$ comprised between 0 and 1.
Maximisation of the cooling power ${I_Q}_{\rm cold}$ regarding to the electrical current $I$ is then obtained for ${I_{\rm MCP} = \alpha T_{\rm cold} / (2R(1-\gamma))}$ and the maximum cooling power is
\begin{equation}
I_{Q_{\rm cold}}^{(\rm max)} = \frac{\alpha^2 T_{\rm cold}^2 }{4R(1-\gamma)}
\end{equation}
\noindent while the EMCP reads
\begin{equation}\label{EMCP}
\epsilon_{\rm MCP} = \frac{\epsilon_{\rm C}}{2 + \epsilon_{\rm C}/(1-\gamma)}
\end{equation}
\noindent 
The above expression of $\epsilon_{\rm MCP}$ is the genuine analogue to the Schmiedl-Seifert efficiency in the refrigerator regime. Note that with the assumption that the temperature difference is negligible compared to the mean temperature, i.e., $\Delta T \ll T_{\rm cold} $, we have $\epsilon_{\rm C} \gg 1$ and an approximate value for $\epsilon_{\rm MCP}$ reads:
\begin{equation}
\epsilon_{\rm MCP} \approx 1 - \gamma
\end{equation}

\section{Discussion and conclusion}
As $\gamma$ varies, the efficiency $\epsilon_{\rm MCP}$ varies between 0 and 1. The EMCP is thus quite small compared to the maximum cooling efficency $\epsilon_{\rm C}$: in the most favourable case one has to provide as much power as heat flux extracted from the cold reservoir to obtain the condition for maximum cooling. This most favourable configuration is reached for $\gamma = 0$, i.e., when all the dissipated heat is released to the cold thermal reservoir, while the EMCP vanishes for $\gamma = 1$, i.e., when all the dissipated heat is released to the hot thermal reservoir. This result seems at first glance counter-intuitive since reinjecting heat where it is extracted from appears as detrimental for the refrigerator operation. However, this dissipated heat acts as a feedback that allows to obtain a maximum for the cooling power; this explains why $\gamma = 0$ yields the highest EMCP for exoreversible engines. It is interesting to note that if $\gamma \rightarrow 1$, $I_{Q_{\rm cold}}^{(\rm max)} \rightarrow \infty$, which is a situation that corresponds to an infinite electrical current, hence a situation where the cooling power does not possess a maximum any longer. A similar behaviour has already been noticed for the $\chi$-criterion optimisation of a strongly dissymetric refrigerator in Ref.~\cite{Izumida2012}.

When theoretical values for the EMCP obtained from Eq.~(\ref{EMCP}) are compared to the practical values used by de Tom\'as and coworkers in Ref.~\cite{deTomas2012}, one immediately notices that the the former are far below the latter. This discrepancy mainly reflects the fact that the real-life working conditions do not correspond to a maximum cooling power but rather to a trade-off between cooling power and cooling efficiency. The exoreversible engine operating in the strong coupling regime may not be the most appropriate way to describe real engines either, since idealisations on both thermal contacts and heat leaks may no longer constitute reasonnable assumptions.

In summary, we have highlighted the reasons of the impossibility to obtain an equivalent to the Curzon-Ahlborn efficiency for the refrigerators: the endoreversible engine with which $\eta_{\rm CA}$ is associated, does not allow a maximum for the cooling power. We have demonstrated that such a maximum exists for exoreversible engines, in which case an expression for the EMCP, the only genuine counterpart of the efficiency at maximum power, has been derived. It does not correspond to the Curzon-Ahlborn efficiency $\eta_{\rm CA}$ but rather to Schmiedl-Seifert efficiency $\eta_{\rm SS}$.  

\acknowledgments
Y.A. acknowledges financial support from the Minist\`ere de l'Enseignement Sup\'erieur et de la Recherche. We also acknowledge support of the Fonds Unifi\'e Interminist\'eriel 7 (SYSPACTE project) and of the French Agence Nationale de la Recherche (ANR), through the program ``Investissements d'Avenir''(ANR-10-LABX-09-01), LabEx EMC$^3$.

\end{document}